\documentclass{mn2e}
\usepackage{amsfonts}
\usepackage{amsmath}
\usepackage{graphicx}

\def\gsim{\;\lower4pt\hbox{${\buildrel\displaystyle >\over\sim}$}\;}
\def\lsim{\;\lower4pt\hbox{${\buildrel\displaystyle <\over\sim}$}\;}
\def\grls{\;\lower4pt\hbox{${\buildrel\displaystyle >\over <}$}\;}

\begin{document}

\title[Solar CME Periodicities]
{Periodicities in Solar Coronal Mass Ejections}

\author[Lou, Wang, Fan, Wang, Wang]
{ Yu-Qing Lou$^{1,2,3}$, Yu-Ming Wang$^{1,4}$, Zuhui Fan$^{1,5}$,
Shui Wang$^{1,4}$,
JingXiu Wang$^{1,5}$\\
$^1$National Astronomical Observatories, Chinese Academy of
Sciences, A20, Datun Road,
Beijing, 100012 China\\
$^2$Physics Department, Tsinghua Center for Astrophysics,
Tsinghua University, Beijing 100084, China\\
$^3$Department of Astronomy and Astrophysics, The University of
Chicago,
Chicago, Illinois 60637 USA; lou@oddjob.uchicago.edu\\
$^4$University of Science and Technology
of China, Hefei, Anhui, 230026, China\\
$^5$Beijing Astrophysical Center and
            Department of Astronomy, CAS-PKU,
            Peking University,
            Beijing 100871 China.  }
%

\date{Accepted ....
      Received ...;
      in original form ...}
\pagerange{\pageref{firstpage}--\pageref{lastpage}}

\pubyear{2003}


\maketitle

\label{firstpage}

\begin{abstract}
Mid-term quasi-periodicities in solar coronal mass ejections
(CMEs) during the most recent solar maximum cycle 23 are reported
here for the first time using the four-year data (February 5, 1999
to February 10, 2003) of the {\it Large Angle Spectrometric
Coronagraph (LASCO)} onboard the {\it Solar and Heliospheric
Observatory (SOHO)}. In parallel, mid-term quasi-periodicities in
solar X-ray flares (class $>$M5.0) from the {\it Geosynchronous
Operational Environment Satellites (GOES)} and in daily averages
of Ap index for geomagnetic disturbances from the {\it World Data
Center (WDC)} at the {\it International Association for
Geomagnetism and Aeronomy (IAGA)} are also examined for the same
four-year time span.
By Fourier power spectral analyses, the CME data appears to
contain significant power peaks at periods of $\sim 358\pm 38$,
$\sim 272\pm 26$, $\sim 196\pm 13$ days
and so forth, while except for the $\sim 259\pm 24$-day
period, X-ray solar flares of class $\gsim$M5.0 show the familiar
Rieger-type quasi-periods at $\sim 157\pm 11$, $\sim 122\pm 5$,
$\sim 98\pm 3$ days and shorter ones until $\sim 34\pm 0.5$ days.
In the data of daily averages of Ap index, the two significant
peaks at periods $\sim 273\pm 26$ and $\sim 187\pm 12$ days
(the latter is most prominent) could imply that CMEs (periods at
$\sim 272\pm 26$ and $\sim 196\pm 13$ days)
may be proportionally correlated with quasi-periodic geomagnetic
storm disturbances; at the speculative level, the $\sim 138\pm
6$-day period
might imply that X-ray flares of class $\gsim$M5.0 (period at
$\sim 157\pm 11$ days) may drive certain types of geomagnetic
disturbances; and the $\sim 28\pm 0.2$-day periodicity is most
likely caused by recurrent high-speed solar winds at the Earth's
magnetosphere. For the same three data sets, we further perform
Morlet wavelet analysis to derive period-time contours and
identify wavelet power peaks and timescales at the 99 percent
confidence level for comparisons. Several conceptual aspects of
possible equatorially trapped Rossby-type waves at and beneath the
solar photosphere are discussed.
\end{abstract}


\section{Introduction}

Mid-term quasi-periodicities (one to several months or longer) in
various diagnostics of solar flare activities and sunspot numbers
or areas etc. during a few years around the solar maximum phase
have been extensively searched for and monitored at many
electromagnetic wavelengths (Rieger et al. 1984; Kiplinger et al.
1984; Dennis 1985; Ichimoto et al. 1985; Delache et al. 1985;
Bogart \& Bai 1985; Bai \& Sturrock 1987; Oliver et al. 1988;
Ribes et al. 1987;  Lean \& Brueckner 1989; \"Ozg\"u\c{c} \&
Ata\c{c} 1989; Lean 1990; Carbonell \& Ballester 1990; Dr\"oge et
al. 1990; Pap et al. 1990; Kile \& Cliver 1991; Verma et al. 1992;
Ballester et al. 1999; Cane, Richardson, \& von Rosenvinge 1998).
These activities of observational research were triggered by the
landmark discovery by Rieger et al. (1984) of a $\sim 154$-day
quasi-period in solar $\gamma$-ray flare rates registered by the
{\it Gamma-Ray Spectrometer (GRS)} onboard the {\it Solar Maximum
Mission (SMM)} two decades ago. Besides this oft-quoted
quasi-period of $\sim$150-160 days, there are other notable
quasi-periods around $\sim$128, $\sim 102$, $\sim 78$, and $\sim
51$ days during the maxima of different solar cycles (Dennis 1985;
Bai 1992; Bai \& Sturrock 1991)
from various data sets. Empirically, these quasi-periods seem to
hint at quasi-subharmonics of equatorial solar rotation period
(e.g., Sturrock \& Bai 1992; Bai \& Sturrock 1993).

As to the plausible physical origin of such quasi-subharmonics
with periods {\it longer} than the solar rotation period, we
proposed recently (Lou 2000a, b) that, in the statistical sense,
the existence of equatorially trapped large-scale (scales
comparable to $R_{\odot}$) Rossby-type waves (Rossby et al. 1939;
Papaloizou \& Pringle 1978; Provost et al. 1981; Saio 1982; Wolff
\& Blizard 1986; Wolff 1998) may be ultimately responsible for
quasi-periodically modulating or triggering smaller-scale magnetic
catastrophes in solar active regions. In this scenario, solar
magnetic active regions become, recurrently, vulnerable to
catastrophes via statistical accumulations of magnetic stresses
and energies through magneto-convective turbulence.

In terms of specific observational diagnostics, we suggested (Lou
2000b) that the photospheric magnetic flux emergence might be
triggered via magnetohydrodynamic (MHD) buoyance instabilities
(Parker 1955) modulated by equatorially trapped Rossby-type waves
owing to the presence of large-scale subphotospheric magnetic flux
(Gilman 1969; Lou 1987; see Ballester et al. 2002 for most recent
emerging magnetic flux observations), that large-scale coherent
velocity patterns or ``cells" may exist over the solar surface
(see Beck, Duvall, \& Scherrer 1998 and Ulrich 2001), and that
more precise measurements of solar surface elevation may reveal
$\lsim $0.1'' variations (see Kuhn et al. 2000 for much smaller
100-m high ``hills" with an azimuthal separation of $[8.7\pm
0.6]\times 10^4$km ). As Rossby-type waves involve fairly slow and
large-scale vortical oscillatory disturbances with periods longer
than the solar rotation period, we speculated (Lou 2000a, b) that
an uninterrupted time sequence of large-scale coronal mass
ejection (CME) events
might contain similar mid-term quasi-periodicities. The {\it
LASCO/SOHO} global observations of solar CMEs in space offer an
unprecedented and unique opportunity for such an investigation.
Uninterrupted {\it LASCO/SOHO} CME data is now available for four
years (February 5, 1999 to February 10, 2003).

Meanwhile, we examine possible mid-term quasi-periodicities for
powerful X-ray solar flares of class $>$M5.0 and daily averages of
Ap index for geomagnetic storm disturbances for several reasons.
First, as a proxy of $\gamma-$ray flares (Rieger et al. 1984),
strong X-ray flares are known to contain Rieger-type periodicities
during past solar maxima. We would like to reconfirm this feature
and provide a test of our current data analysis at the same time.
Secondly, strong X-ray flares might correlate with global CMEs in
some statistical sense. Thirdly, in terms of both space weather
and solar terrestrial interactions, we would like to establish
possible correlations among CMEs, intense X-ray flares, and
geomagnetic disturbances characterized by daily averages of Ap
index. Systematic investigations on quasi-periodicities may reveal
such correlations in a unique manner.

For all three data sets, we shall first use Fourier analysis to
identify possible quasi-periodicities. To complement the Fourier
spectra, we further carry out Morlet wavelet analyses to derive
contour plots in period-time domains for the same three time
series of CME, X-ray solar flares, and geomagnetic Ap index data,
respectively. With a few exceptions, major peaks identified in the
wavelet analyses are roughly consistent with major peaks in the
Fourier power spectral analyses. Additional information of when
certain quasi-periodicities occur or recur in different data sets
may provide valuable clues for possible physical connections.

In broader astrophysical contexts, we note in passing that global
MHD tidal waves including Alfv\'en-Rossby type waves (Lou 1987,
2000b, 2001) may give rise to valuable periodic or quasi-periodic
signatures from magnetized compact stars such as neutron stars and
white dwarfs that are covered by corotating thin yet dense
``plasma oceans" involving extremely strong magnetic fields
ranging from $10^7$ G to $10^{15}$ G. It is our desire to use the
Sun as an astrophysical laboratory to search for, to identify, and
to understand stellar Rossby-type waves.

\begin{figure}
\begin{center}
\caption{ (a) Daily counts of CME events from LASCO for the
four-year time span shown on top. (b) Fourier power spectrum of
the CME data. (c) Comparison of statistical probability for power
larger than a preset value between the CME data sequence (asterisk
$*$) and an artificially yet sufficiently randomized CME data
sequence ($+$).}
\end{center}
\end{figure}

\section{Data Analysis and Results\\
on Quasi-Periodicities}

The time sequence of daily CME counts was derived from a
preliminary list of {\it LASCO/SOHO} compiled by the Naval
Research Laboratory (NRL) at the website
lasco-www.nrl.navy.mil/cmelist.html.
The solar X-ray flare data were taken from
the {\it GOES} at the website
www.ngdc.noaa.gov/stp/SOLAR/solar.html and from the website
www.sec.noaa.gov/Data/solar.html.
The data of the daily averaged Ap index for geomagnetic storm
disturbances were obtained from the {\it WDC}
of the {\it IAGA}
at the website swdcdb.kugi.kyoto-u.ac.jp/wdc/Sec3.html.

\subsection{Fourier Power Spectral Analyses}

The uninterrupted data for daily counts of CMEs from {\it
LASCO/SOHO} are displayed in Fig. 1a. Prior to Feb. 5, 1999, there
were several unfortunate interruptions of {\it SOHO} operations
and those earlier CME data are therefore excluded. Fig. 1b is the
Fourier power spectrum of the data shown in Fig. 1a. In spite of
the dominance (Fig. 1a and 1b), peak A at $\sim$1101-day period is
not reliable as this period is not much less than the total data
length of 1467 days (edge effects). While being preliminary, the
next three significant peaks B, C, and D at quasi-periods of $\sim
358\pm 38$, $\sim 272\pm 26$, and $\sim 196\pm 13$ days,
respectively, are longer than the usual Rieger-type quasi-periods
(Rieger et al. 1984; Dennis 1985; Sturrock \& Bai 1992) in the
flare-related data (see Table 1 for significant CME power peaks
identified). To estimate possible errors in periods of identified
power peaks, we take half of the full width at half maximum (FWHM)
about a power peak in Fig. 1b as the angular frequency variation
$\Delta\omega$. Then the period variation is given by $\Delta
P=-P\Delta\omega/\omega$. That is, $P\pm\Delta P$ would give an
estimated period variation range for an identified power peak. In
the same manner, we have obtained estimates for power spectra of
X-ray flares (class $>$M5.0) and of daily averages of Ap index
(see Table 1 for details). By Table 1, $\Delta P$ increases with
increasing $P$ in general.

\begin{figure}
\begin{center}
\caption{The same panel arrangement as in Fig. 1 but for the X-ray
solar flare data (class $>$M5.0) from GOES during the same time
span indicated on top.}
\end{center}
\end{figure}

\begin{figure}
\begin{center}
\caption{The same panel arrangement as in Fig. 1 but for the daily
averaged Ap index of geomagnetic disturbances from the WDC/IAGA
during the same time span indicated on top.}
\end{center}
\end{figure}

\begin{table*}
\caption{\hskip 2.5cm List of quasi-periodicities (Data sets from
February 5, 1999 to February 10, 2003)}
\begin{tabular}{cc|cccccccccccc}
\hline Item&\multicolumn{12}{c}{Identified Quasi-Periods
by Spectral Power Peaks in Unit of Days} \\
&\hbox{}&A&B&C&D&E&F&G&H&I&J&K&L \\
\hline CME&$P$&1101.22&358.33&271.99&195.88&110.80&100.00&66.25
&60.64&57.26&35.85&33.49&20.62 \\
&$\Delta P$&671.48&37.68&26.22&13.38&3.26
&2.88&1.31&0.90&1.00&0.40&0.39&0.17 \\
\hline
Flares&$P$&1505.00&259.48&156.77&122.19&98.15&67.69&63.73&42.16&38.71&33.53&& \\
&$\Delta P$&589.46&24.23&10.89&4.88&3.25&2.18&2.34&0.78&0.55&0.52&& \\
\hline Ap&$P$&940.63&364.11&272.81&187.34&138.07&91.49&60.64
&27.80&27.08&24.69&24.00&23.29 \\
&$\Delta P$&249.85&38.91&25.55&12.05
&6.33&5.19&2.38&0.22&0.21&0.17&0.25&0.14 \\
\hline
\end{tabular}\\[4pt]
{\small All original data were taken from February 5, 1999 to
February 10, 2003. Significant power peaks based on Fourier
analyses are identified and summarized here.
We have periodicity$=P\pm\Delta P$ and $\Delta
P=-P\Delta\omega/\omega$ where $\Delta\omega$ is a half of the
full width at half maximum (FWHM) about a power peak. See Figs. 1,
2, and 3 in the text for CME, X-ray solar flares, and Ap index,
respectively.}
\end{table*}
%

For time series of the daily CME data shown in Fig. 1, we took the
following steps to estimate the statistical significance of
Fourier spectral power peaks (e.g., Delache et al. 1985). First,
the daily CME data is picked at random within the time sequence
and rearranged several to more than ten times to form an
artificially randomized sequence (similar to the process of
shuffling cards) with much reduced coherent periodic signals if
any. For such a white noise sequence of independent Gaussian
distributions with variance $\sigma^2$, a Fourier transform would
yield a power spectrum characterized by a probability distribution
$p(w)=\exp [-w/(2\sigma^2)]$ for spectral powers greater than a
preset value $w$. The approximate straight line of symbol $+$ in
Fig. 1c represents the natural logarithm of $p(w)$, $\ln [p(w)]$,
versus $w$ for an artificially randomized CME data sequence with a
variance of
$\sigma^2\cong 1.46\times 10^{-3}$ ($\sigma\cong 3.8\times
10^{-2}$) estimated from the slope of this line. The relevant
$\sigma$-levels of {\it fluctuation amplitudes} are then shown
(viz., the two horizontal dashed lines in Fig. 1b corresponding to
3-$\sigma$ and 4-$\sigma$ levels as well as a sequence of numeral
marks on top of Fig. 1c). Next, the natural logarithm of
probability for spectral powers greater than a preset value $w$ is
estimated from the power spectrum of the actual daily CME counts
(symbol $*$ in Fig. 1c). A comparison of the two curves in Fig. 1c
clearly indicates that for powers greater than the
$3\sigma$-level, the CME data starts to depart from random noises.
Power peaks B through K in Fig. 1c are regarded as significant.
Periods of peaks J and K are $\sim 36\pm 0.4$ and $\sim 33\pm 0.4$
days, respectively. We ran the same test procedure for the other
two data sets and obtained qualitatively similar properties. That
is, for both Ap index and X-ray solar flares of class $\gsim$M5.0
data sets examined below, deviations from random noises are also
notable for power peaks $\gsim 3\sigma$-level (see Figs. 2c and
3c).

It should be noted that such a procedure can be executed whatever
the count distribution of the original data sequence may be. In
fact, we have obtained data count distributions about the
respective means for all three data sequences (i.e., CMEs, X-ray
flares, and Ap index). It turns out that both CME and X-ray flare
data more or less follow a Poisson distribution (that is,
$P_\mu(\nu)=\exp(-\mu)\mu^{\nu}/\nu !$ where $\mu$ is the mean
count rate and $\nu$ is the actual count rate; see e.g., Taylor
1981), while the Ap data deviate significantly away from a Poisson
distribution (not surprisingly).

Entirely parallel to the panel arrangement of Fig. 1, Figure 2
presents a similar analysis on daily counts of X-ray solar flares
of class $\gsim$M5.0 from the {\it GOES}. By Fig. 2c, the
estimated variance is
$\sigma^2\cong 3.52\times 10^{-5}$ ($\sigma\cong 5.9\times
10^{-3}$). As a confirmation, the X-ray flare data in Fig. 2a does
contain the familiar Rieger or Rieger-type quasi-periodicities
$\lsim 157\pm 11$ days (see Fig. 2b and Table 1 for power peaks
identified in the solar X-ray flare data) which do not have
significant counterparts in the CME power spectrum except for
periods of $\sim 34\pm 0.5$, $\sim 39\pm 0.6$, and $\sim 98\pm 3$
days (Fig. 1b and Table 1). As we deal with the most powerful
X-ray flares, this appears to be consistent, in trend, with the
fact that Rieger-type quasi-periodicities were first discovered
(Rieger et al. 1984) in $\gamma-$ray flares by the {\it SMM/GRS}
followed by detections in soft X-ray flares (Kiplinger et al.
1984). Note that peak B of $\sim 259\pm 24$-day period appears
significant enough ($> 4\sigma$ level) and may be related to the
$\sim 272\pm 26$-day period in the CME data. For reasons noted
earlier, peak A at $1505\pm 589$-day period again might not be
reliable even though its apparent confidence level is merely
slightly $\lsim 95\%$.

Similarly, Figure 3 contains the relevant information of the daily
averaged Ap index for geomagnetic storm disturbances, with an
estimated variance
$\sigma^2\cong 0.08$ ($\sigma\cong 0.28$). Power peaks C and D in
Fig. 3b at periods $\sim 273\pm 26$ and $\sim 187\pm 12$ days may
correspond to the two peaks at periods $\sim 272\pm 26$ and $\sim
196\pm 13$ days of the CME data in Fig. 1b. There also exists a
fairly prominent peak H ($>4$-$\sigma$ level) at a $\sim 28\pm
0.2$-day period. The $\sim 138\pm 6$-day period might correspond
to the $\sim 157\pm 11$-day period in X-ray flare data;
admittedly, this correspondence is somewhat weak given the ranges
of error estimates. Should this correlation be real, then there
might exist subclasses of geomagnetic disturbances which are
proportionally correlated with major X-ray solar flares (class
$>$M5.0). As already suspected, the $941\pm 250$-day period may
not be reliable due to edge effects. In sharp contrast, by a
visual inspection of Fig 3a and 3b, the most dominant power peak D
at period of $\sim 187\pm 12$ days for Ap index should be
physically real.
In Fig 3, the prominent peak H around $\sim 28\pm 0.2$-day period
in the daily averaged Ap index is physically identified with the
solar rotation that recurrently brings high-speed solar wind
streams from low-latitude coronal holes towards the Earth's
magnetosphere. Note that this $\sim 28\pm 0.2$-day period is
almost absent in the CME data.

\subsection{Considerations of Statistical Significance}

\begin{figure}
\caption{Period-time contours of Morlet wavelet analysis on the
solar CME data (Feb 5, 1999 to Feb 10, 2003) with $\omega_{0}=6$
and the normalized power. Timescales (periods) of power peaks $A$
through $E$ are summarized in Table 2. The cone of influence (COI)
is beneath the dashed curve.}
\end{figure}

\begin{figure}
\caption{Period-time contours of Morlet wavelet analysis on the
X-ray solar flare data (class $>M5.0$) with $\omega_{0}=6$ and the
normalized power. Timescales (periods) of power peaks $A$ through
$F$ are summarized in Table 2. The cone of influence (COI) is
beneath the dashed curve.}
\end{figure}

\begin{figure}
\caption{Period-time contours of Morlet wavelet analysis on the Ap
index data with $\omega_{0}=6$ and the normalized power.
Timescales (periods) of power peaks $A$ through $D$ are summarized
in Table 2. The cone of influence (COI) is beneath the dashed
curve.}
\end{figure}

Figs. 1c, 2c, and 3c provide a sense of statistical significance
for power peaks identified in the periodograms of CME, X-ray
flare, and daily averaged Ap index data, respectively. Basically,
when a spectral amplitude (square root of a power peak) becomes
higher than the $3\sigma$-level, the statisitics of corresponding
power peaks in the periodograms (asterisks $*$) starts to deviate
from power peaks derived from randomized data series (crosses
$+$). The higher the spectral amplitude, the larger the deviation
in statistics.

Any one of the time series analyzed here contains random
fluctuations that may not necessarily obey Gaussian distribution.
The corresponding power spectrum naturally contains random
variations. For independent daily counts of sufficiently large
number in a data sequence, the corresponding Fourier amplitudes
obey Gaussian distribution by the central limit theorem (e.g., Fan
\& Bardeen 1995). Therefore, the power probability distribution of
$Z\equiv P(\omega)$ being in the interval $z<Z<z+dz$ is
$Pr\{z<Z<z+dz\}=\exp(-z)dz$ where $P(\omega)$ is normalized by
$\sigma^2$. The cumulative distribution function is
$Pr\{Z<z\}=\int_0^{z}\exp(-z')dz'=1-\exp(-z)$. Consequently, the
probability for $Z>z$ is $Pr\{Z>z\}=\exp(-z)$. To search for the
maximum value of power peaks among $N$ independent frequencies
such that $Z$ are independent random variables, the corresponding
probability is
$$
Pr\{Z_{max}>z\}=1-[1-\exp(-z)]^N\ .
$$
By this expression, it follows that a preset threshold $z=z_{0}$
for power peaks among $N$ independent variables is related to the
probability (false alarm probability $p_{0}$) that a peak, being
produced randomly by chance, is greater than $z_{0}$ (Lomb 1976;
Scargle 1982; Ballester et al. 2002).

To further assess the statistical significance of power peaks in
Figs. 1 to 3 in this approach, we estimate the detection threshold
$z_{0}=-ln[1-(1-p_0)^{1/N}]$ in spectral power with a false alarm
probability $p_0$, where $N$ is the number of independent
frequencies over which a power peak is searched for (Scargle
1982). In our case, $N=734$ (a half of the total number of days of
observations); for a false alarm probability $p_{0}=0.01$ ($99\%$
confidence), we have a detection threshold of $z_{0}\cong
11.2\sigma^2$ and for $p_{0}=0.05$ ($95\%$ confidence), we have a
detection threshold of $z_{0}\cong 9.6\sigma^2$. In Fig. 1, peaks
$B-D$, $F$, $G$, $J$, $K$ are above the $99\%$ confidence level;
and peaks $E$, $H$, and $I$ are above the $95\%$ confidence level.
In Fig. 2, peaks $B-E$ and $H-J$ are above the $99\%$ confidence
level. In Fig. 3, peaks $C$, $D$, $H$, $I$, and $J$ are above the
$99\%$; and peaks $B$, $F$, $G$, and $K$ are above the $95\%$
confidence level.

The tentative correlation found between mid-term
quasi-periodicities of the CME and Ap data (Figs. 1 and 3) can be
of physical significance. Besides the cause of recurrent
high-speed solar winds with southward interplanetary magnetic
fields for geomagnetic disturbances, a geomagnetic storm may set
in when a CME or a part of a CME impinges upon the Earth's
magnetosphere (Cane et al. 2000; Wang et al. 2002). \footnote{For
example, Jupiter's hectometric radio emissions and extreme
ultraviolet aurorae were found to be triggered by arrivals of
interplanetary shocks according to simultaneous observations of
the Cassini and Galileo spacecraft (Gurnett et al. 2002).} Among
all solar CMEs, there is certainly a significant fraction that
miss the Earth's magnetosphere owing to their various initial
onset orientations over the solar surface. Statistically, the
probability, roughly proportional to the solid angle subtended by
the Earth's magnetosphere towards the Sun, that CMEs hit the
Earth's magnetosphere is higher for more frequent occurrence of
CMEs. It is in this proportional sense that periodicities in CME
and Ap data should correlate with each other. This is an important
perspective for probing solar-terrestrial interactions and space
weather conditions. This empirical correlation also strengthens
our confidence that the two low-frequency power peaks C and D in
the CME data are likely to be physically real.

The solar X-ray flare $\sim 259\pm 24$-day period is close to the
CME $\sim 272\pm 26$-day period and the Ap $\sim 273\pm 26$-day
period. If not coincidental by chance, this might indicate a
certain causal relation among energetic X-ray solar flares
($\gsim$M5.0) and CMEs. Meanwhile, we caution that it would be
premature to claim the existence of different types of solar
flares associated with global CMEs.


\begin{table}
\caption{\hskip 0.2cm List of power peaks in Morlet wavelet
contours}
\begin{tabular}{c|cccccc}
\hline\\
Item&\multicolumn{6}{|c}{Periods in Days for Contour Peaks}\\
 &A&B&C&D&E&F\\
\hline\\
CMEs&343.0&187.0&102.0&38.3&36.1&\\
X-ray Flares&242.5&144.2&72.1&66.1&39.3&25.5\\
Ap index&288.4&187.0&66.1&51.0&&\\
\hline
\end{tabular}\\
{\small In normalized Morlet wavelet period-time contours of Fig.
4 (CME data), Fig. 5 (X-ray flare data), and Fig. 6 (Ap index
data), we identify power peaks and timescales (periods) at the 99
percent confidence level (1 percent significance), respectively.}
\end{table}

\subsection{Morlet Wavelet Analyses}

As already noted earlier, mid-term quasi-periodicities of various
solar flare related diagnostics may change for maxima of different
solar cycles. It is thus of considerable interest to see whether
such periodicities change within a few years around the maximum
phase of a solar cycle. It is also important to check whether
power peaks of Fourier spectral periods and of wavelet contour
timescales have reasonable correspondence. For correlation studies
of different time series data, it is crucial to know at what times
certain periodicities occur or recur.

For these purposes, we perform wavelet analysis on the three time
series for the CMEs, solar X-ray flares, and Ap index using the
Morlet wavelet function
\begin{equation}
\psi_{0}(\eta)=\pi^{-1/4}e^{i\omega_{0}\eta}e^{-\eta^{2}/2}\
\end{equation}
where $\omega_{0}=6$ is chosen. The Fourier time period $\tau$ and
the Morlet wavelet timescale $s$ are related by
\begin{equation}
\tau=4\pi s/[\omega_0+(2+\omega_0^2)^{1/2}]\
\end{equation}
(see Table 1 of Torrence \& Compo 1998). For our choice of
$\omega_{0}=6$, it follows from equation (2) that $\tau=1.033s$,
namely, Fourier periods and Morlet wavelet timescales are nearly
equal to each other. The wavelet transform suffers from edge
effects at both ends of the time series. This gives rise to a cone
of influence (COI) as indicated by regions below the dashed lines
in Figures 4, 5, and 6. Caused by padding zeroes at the beginning
and at the end of data series, these edge effects usually lead to
a power reduction within the COI. For further detailed technical
information of Morlet wavelet analysis, one may visit the website
at http://paos.colorado.edu/research/wavelets/wavelet2.html.

At the 99 percent confidence level (1 percent significance) and
using the Morlet wavelet function defined by equation (1), contour
plots in period-time (or equivalently, scale-time) domains with
color coding are shown for the CME, X-ray solar flares, and Ap
index data in Figures 4, 5, and 6, respectively, and corresponding
contour peaks are summarized in Table 2.

Comparing Table 1 and Table 2, we find that CME Fourier periods
$B$, $D$, $F$, and $J$ at $\sim 358\pm 38$, $\sim 196\pm 13$,
$\sim 100\pm 3$, and $\sim 36\pm 0.4$ days roughly correspond to
CME Morlet periods $A$, $B$, $C$, and $E$ at $\sim 343$, $\sim
187$, $\sim 102$, and $\sim 36$ days, respectively; X-ray flare
Fourier periods $B$, $C$, $F$, and $I$ at $\sim 259\pm 24$, $\sim
157\pm 11$, $\sim 68\pm 2$, and $\sim 39\pm 0.6$ days roughly
correspond to X-ray flare Morlet periods $A$, $B$, $C$, and $E$ at
$\sim 243$, $\sim 144$, $\sim 66$, and $\sim 39$ days,
respectively; and Ap index Fourier periods $C$, $D$, and $G$ at
$\sim 273\pm 26$, $\sim 187\pm 12$, and $\sim 61\pm 2$ days
roughly correspond to Ap index Morlet periods $A$, $B$, and $C$ at
$\sim 288$, $\sim 187$, and $\sim 66$ days, respectively.

The most prominent match is the $\sim 187$-day period in both CME
and Ap index data (this quasi-periodicity is also present in
corresponding Fourier power spectra). As already noted earlier, we
suspected that this Ap index periodicity was driven by periodic
CMEs.
%
%
These comparisons suggest that CMEs and Ap index for geomagnetic
disturbances may proportionally correlate with each other in both
periods and time.  We categorically emphasize that power peaks in
Ap index data around $181-187$ days are so prominent in Figs. 3a,
3b, and 6, and that they should be physically real for whatever
origin. Here, the evidence lends support for the CME-driven
scenario in the statistical sense.

Wavelet power peak in C$-$D range of Fig. 6 (Ap index) and wavelet
power peak in C$-$D of Fig. 5 (X-ray flares) have comparable
timescales of $\sim 66$ days. They occur around the same time and
may be interpreted as flare-driven periodicities in Ap index data.
There is also a match around periodicity of $\sim 39$ days in both
X-ray flare and Ap index data. However, there is a time difference
for the two relevant power peaks in CME and X-ray flare data. From
such information alone, correlations between solar X-ray flares
and CMEs at these timescales are not immediately clear.

\subsection{A Test of Incomplete X-Ray Flare Data}

Such CME mid-term quasi-periodicities (present in both Fourier
spectrum and Morlet wavelet analyses), if confirmed by further
independent observations, offer novel diagnostics for probing and
understanding the physical origin of CMEs (e.g., Hundhausen 1999;
Low 2001). Up to this point, one major question arising from our
comparative data analysis is that there are mid-term
quasi-periodicities in the {\it LASCO/SOHO} data for CMEs but
these periodicities, except for a few, are largely independent of
the Rieger-type quasi-periodicities detected in various phenomena
associated with energetic X-ray solar flares.

While the {\it SOHO/LASCO} data of CMEs including all halo events
is the most complete one, the X-ray solar flare data of the {\it
GOES} suffer incompleteness because one cannot see the other half
of the Sun at any given moment. Might this be the cause of
differences in some quasi-periodicities seen in CMEs and solar
X-ray flares? We designed a simple test to address this issue with
the assumption that a strong X-ray solar flare occurs at one
footpoint of a CME. In this scenario, a CME may be related to one
flare at most. For example, we may observe a CME without seeing an
intense X-ray flare perhaps because one footpoint of the CME is
located on the frontside but another is located on the backside of
the Sun and the corresponding flare happens at the footpoint on
the backside. By all combinations, $M$ CMEs per day may correspond
to $M+1$ (0,1,...,M) possible X-ray flare counts on the frontside
of the Sun. Based on the CME data, we then rely on a random number
generator to select one of the possibilities as observable flare
events per day. The key features of the resulting power spectrum
thus obtained do not change significantly for arbitrary trials.
This test seems to indicate that should a more complete data set
of solar flares be available, the major quasi-periodic features in
the power spectrum would remain.

\section{Mid-Term Quasi-Periodicities}

On the basis of our data analyses reported here, it appears that,
with high probability, mid-term quasi-periodicities do exist in
association with diagnostics of various solar activities (CMEs and
X-ray solar flares) as well as geomagnetic disturbances (Ap
index). Some of the overlapping quasi-periodicities might reveal
underlying physical causes. We shall discuss qualitatively
plausible physical connections.

\subsection{Quasi-Periodicities in Flux Emergence}

Regarding the theoretical suggestion of equatorially trapped
Rossby waves (Lou 2000a, b) for Rieger-type periodicities in
flare-related diagnostics, we note that intuitively, solar flare
rate and emergence of complex magnetic regions of sunspots or
sunspot groups should relate to each other in a statistically
correlated manner (Oliver et al. 1998; Ballester et al. 1999,
2002). Empirically, nearly all solar flares are found close to
sunspots or sunspot groups and often seen as brightenings of the
pre-existing plages; the observed quasi-periodicities (Bai 1992;
Oliver et al. 1998) in sunspot areas or number of sunspot groups
do indeed correlate with those of flare occurrence rates. Is this
merely a coincidence? Or, does this actually hint at an underlying
global mechanism? We here discuss two seemingly independent
aspects, namely, the emergence of magnetic fluxes and the
occurrence of solar flares, even though the two aspects should in
some sense be related to each other empirically.

Whether magnetic fields are generated deep in the Sun's radiative
interior (e.g., Gough \& McIntyre 1998) or at the bottom of the
solar convection zone (e.g., Rosner \& Weiss 1985), some of these
magnetic fluxes must somehow float upward through the convection
zone and eventually break through the thin photosphere by magnetic
buoyancy (Parker 1955, 1979) to buckle upwards above the solar
surface, forming sunspot pairs or groups. In the absence of a
global quasi-periodic modulation mechanism, such process of
magnetic flux emergence should be completely random. The observed
quasi-periodicities in magnetic flux emergence would be consistent
with (though by no means necessary) the presence of large-scale
Rossby-type waves of comparable periodic timescales. Generally
speaking, equatorially trapped Rossby-type waves (Lou 2000a, b)
can dynamically affect subsurface as well as emerged large-scale
magnetic fields (e.g., magnetic active regions) through squeezing,
twisting, stretching and vortical motions, even though spatial
scales of (emerged) individual sunspots or sunspot groups are
relatively small.\footnote{The process leading to the formation of
such intense magnetic structures is nonlinear and complicated
(Parker 1955).} More specifically, such equatorially trapped
Rossby-type waves may tip off vulnerable regions for the onset of
magnetic buoyancy instabilities via dynamic couplings. This
plausibly links quasi-periodicities detected in sunspot areas or
number of sunspot groups (Lean 1990; Oliver et al. 1998; Ballester
et al. 1999, 2002) with equatorially trapped Rossby-type waves.

\subsection{Quasi-Periodicities in Flare Diagnostics}

The occurrence of solar flares involves sudden releases or bursts
of considerable magnetic energies accumulated over a certain
period of time (e.g., Parker 1979). It is almost impossible to
predict the location and time of a solar flare.\footnote{This
situation is analogous to hurricanes, earthquakes, and tornadoes
in that vulnerable environments for their occurrences can be
identified with relative ease while their specific occurrence
location and time are very difficult to pinpoint or predict.} As
magnetic fields associated with sunspots or sunspot groups are
strong in strengths and complicated in structures, it is quite
natural to somehow have a significant amount of magnetic energy
stored in special ways as magnetic fields gradually evolve in time
(Low \& Lou 1990; Lou 1992). In contrast to nanoflares (Parker
1994) which frequently release magnetic energies on smaller scales
(e.g., X-ray bright points) for heating the lower corona, this
accumulation (or storage) stage of magnetic energies is necessary
for flare phenomena in general otherwise solar flares would not be
so violent and explosive (e.g., Zirin 1988). At critical moments,
small disturbances of whatever origin may {\it trigger} magnetic
avalanches, leading to productions of energetic particles and
wide-spectrum electromagnetic radiations.
Only in this sense, a sustained passage of Rossby-type waves can
increase {\it statistically} the chance of flare occurrence in
preexisting vulnerable magnetic complexes and thus impose
quasi-periodicities to various diagnostics associated with X-ray
solar flares.

\subsection{Dynamic Feedback Cycle}

Having argued heuristically about possible roles of equatorially
trapped Rossby-type waves in triggering the emergence of magnetic
fluxes and the onset of solar flares in a statistically correlated
manner, we need to address the pressing question of Rossby-type
wave excitation (Lou 2000a, b). Whether such waves (with much
weaker magnitudes) exist or not during the solar minimum phase is
not presently known. Nevertheless, preceding discussions require
their presence as a large-scale coordinating mechanism during the
solar maximum phase.

As a solar flare or a CME goes off, a fraction of their energies
will back react on the solar photosphere and appear in the form of
traveling disturbances. Collectively, a fraction of total flare or
CME energies may provide a necessary energy source for generating
and sustaining equatorially trapped Rossby-type waves during a few
years around solar maxima. In this scenario, a {\it dynamic
feedback cycle} may be established to sustain quasi-periodicities
in magnetic flux emergence and flare activities. Failure of
sustaining such a feedback cycle leads only to {\it occasional}
periodicities in flare activities (see Fig. 5). This may explain
why during maxima of some solar cycles, the 150-day
quasi-periodicity becomes so prominent (Oliver et al. 1998) while
in others it could be almost absent with yet different
periodicities more overwhelming (cf. Bai 1992; Bai \& Sturrock
1991, 1993; Sturrock \& Bai 1992) --- the dominant periodicity can
be different during solar maxima of different solar cycles (e.g.,
Bai \& Sturrock 1991; Bai 1992) and for different diagnostics
(Oliver et al. 1998; Ballester et al. 1999, 2002). Fig. 5 clearly
shows time variations of periodicities even within a few years of
the recent solar maximum phase (cycle 23). This proposed feedback
scenario leaves considerable rooms for possible correlations among
intensities/rates of solar flares, selection of periodicities, and
distribution of sunspot groups on both northern and southern sides
of the solar equator during solar maxima and therefore also
requires further explorations.

\subsection{Quasi-Periodicities in CME Diagnostics}

CMEs from the solar limb have been routinely monitored by various
coronagraph observations (cf. Hundhausen 1999 for SMM results).
With the ongoing Solar and Heliospheric Observatory (SOHO) mission
in space, even relatively weak CME halo events\footnote{The
strengths of these halo CMEs are not necessarily weak. They are
not as easy to detect merely because they originate somewhere
within the solar disk.} can now be readily detected as well. By
combining SOHO and ground-based observations, it is possible to
construct a fairly complete time sequence for CMEs occurrence.

CMEs represent a major class of large-scale solar activities
involving magnetic field and perhaps sub-photospheric magnetic
flux emergence. Physically, CMEs are characterized by large-scale
eruptive mass losses from the Sun into the solar wind (Hundhausen
1999) and are thought to be caused by a sudden loss of magnetic
equilibria (Low 1990) as photospheric magnetic field structures
evolve gradually in time. The primary energetics of CMEs is
believed to be magnetic in nature. At a critical stage of
large-scale magnetic structure evolution, a CME may be {\it
triggered} by disturbances. As a systematic and quasi-periodic
source of large-scale disturbances at the photospheric level,
equatorially trapped Rossby-type waves may modulate CMEs in a
periodic yet statistical manner.
%

\section{Application of Rossby-Type Waves}

Our comparative observational research is motivated by the known
mid-term quasi-periodicities, namely, Rieger-type periodicities,
in various solar flare related diagnostics and our analysis here
on X-ray flares (class $>M5.0$) during the recent solar maximum of
cycle 23 confirms once again the existence of such Rieger-type
periodicities. What is then the underlying physics for such
``discrete quasi-periods" arranged roughly like subharmonics and
significantly longer than the solar rotation period if they are
indeed real? Solely from the perspective of period matchings, it
appears that solar Rossby-type waves are the best bet for such
discrete subharmonic quasi-periodicities, although many questions
remain to be addressed for this proposal (Lou 2000a, b).

Now given the global nature as well as randomness of CME events,
how does one infer the underlying physics for mid-term
quasi-periodicities (see Fig. 1 and Table 1) longer than those of
the Rieger type? The reality of such mid-term quasi-periodicities
for CMEs appears to be indirectly supported by the presence of
similar quasi-periodicities in the daily average of Ap index (see
Fig. 3 and Table 1). The basic rationale for such a remote
connection is that quasi-periodic CMEs should lead to
quasi-periodic geomagnetic disturbances when a fraction of CMEs
encounter the Earth's magnetosphere. It is certainly possible to
match the identified quasi-periods of CMEs with periods of
Rossby-type waves (Lou 2000a, b) as we did for Rieger-type
periodicities. However, as shown in Table 1, longer periods have
larger error ranges comparable to the equatorial solar rotation
period, that is, the accuracy is insufficient to warrant a
one-to-one correspondence without ambiguities.

With these qualifications and limitations in mind, we shall
nevertheless discuss several empirical, intuitive and conceptual
aspects of mid-term quasi-periodicities in reference to
equatorially trapped Rossby-type waves.

For Rieger-type periodicities, we may estimate the quasi-periods
of equatorially trapped Rossby-type waves by taking $m=12, 10, 8,
6$ and 4 with $n=1$ or $2$ in expressions (13) and (15) of Lou
(2000b). We suspect that the spatial distribution of sunspot
groups or active regions along the equatorial zone might be one
important factor in the selection of the two integers $m$
($k_x\equiv m/R_{\odot}$ is the azimuthal wavenumber) and $n$
(number of nodes for a Rossby wave function along a longitude)
during the solar maximum phase. By definition, a solar minimum is
characterized by a fewer number of sunspots and thus less
activities. During a solar maximum in contrast, radio images of
the Very Large Array (VLA) as well as X-ray images from Yohkoh
reveal two quasi parallel chains of 5 or 6 emission-intense
``blobs" from sunspot groups or magnetic complexes on both
northern and southern hemispheres.\footnote{In some cases, the
separation of different ``blobs" may involve certain ambiguities.
Sometimes, one might hesitate as whether to count a ``blob" at the
limb or not. For the present purpose, we mainly focus on the basic
concept of such mode selection.} As these ``blobs" are sites of
frequent flare occurrence, excited equatorially trapped
Rossby-type waves are likely to concentrate their powers in the
relevant $m$ which roughly corresponds to the number of ``blobs"
around the Sun. For $\sim$5 or 6 active centers in the solar disk
facing us, $m$ would then be $\sim$10 or 12 around the
circumference. In turn, Rossby-type wave disturbances will
statistically trigger more flares in active regions. In essence,
this is part of the dynamic feedback cycle outlined earlier.

Likewise, the gross symmetry for the numbers of active regions
across the equator would correspond to $n=2$. Thus, equatorially
trapped Rossby-type waves with $n=2$ might be favorably excited
given the presence of two belts of sunspot groups across the
equator. It should be noted that for wave modes with even $n$, the
north and south are symmetric with respect to the equator; and for
wave modes with odd $n$, the north and south are antisymmetric
with respect to the equator. Of course, for a mixture of
Rossby-type wave modes with both odd and even $n$, it is not
possible to identify a symmetry with respect to the equator in the
strict sense.

Observationally, the distribution of active regions is sometimes
not grossly symmetric with respect to the equator during a certain
epoch of a solar maximum phase and this situation may correspond
to odd values of $n$. This might be the cause of north/south
asymmetries sometimes observed in periodicities of flare rates
(e.g., Bai 1987). One should keep in mind that for $n=1$ or $2$
and $m\geq 4$, the angular wave pattern speeds
$\omega_p\equiv\omega/m$ of equatorially trapped Rossby-type waves
are approximately proportional to $\sim 2\Omega_{\odot}/m^2$. The
larger the value of $m$, the slower the angular wave pattern speed
$\omega_p$. By including effects of surface elevation (see
equations 1-3 of Lou 2000), periods of Rossby-type wave crest
bumping into the next adjacent active region are
$2\pi/|m\omega_p|$ as specifically given by expressions (13) and
(15) of Lou (2000b), namely,
\begin{equation}
P_r\cong P_{\odot}\lbrace |m|/2+(2n+1)\Omega_{\odot}
R_{\odot}/[|m|(gD)^{1/2}]\rbrace\
\end{equation}
for Rossby-wave periods, and
\begin{equation}
P_{r-p}\cong\lbrace |m|+[{m^2+8\Omega_{\odot}R_{\odot}
/(gD)^{1/2}}]^{1/2}\rbrace P_{\odot}/4\
\end{equation}
for Rossby-Poincar\'e-wave periods, respectively, where
$P_{\odot}\equiv 2\pi/\Omega_{\odot}$ is the Sun's sidereal
rotation period, $\Omega_{\odot}R_{\odot}\sim 2\hbox{ km s}^{-1}$
is the equatorial solar rotation speed and $(gD)^{1/2}$ is the
solar surface gravity wave speed with $g\sim 2.7\times 10^4\hbox{
cm s}^{-2}$ being the solar surface gravity and $D$ (of a few
hundred kilometers) being an effective thickness of the
photospheric layer.
%

The solar photosphere is magnetized with well-known
inhomogeneities on smaller scales such as concentrations of
intense magnetic fibrils ($\sim 10^3$ G) along boundaries of
supergranules whose typical diameters are on the order of $\sim
3\times 10^9$ cm and appearances of sunspots or sunspot groups (of
sizes comparable to that of a typical supergranule) with magnetic
field strengths up to $3 - 4\times 10^3$ G. Subsurface horizontal
magnetic fields $B_h$ (e.g., Babcock \& Babcock 1955) may well be
patchy on large scales of the order of $L\sim R_{\odot}$ and might
be fairly strong deep in the solar interior (e.g., Gough \&
McIntyre 1998). However, as the gas density $\rho$ increases
rapidly with depth by gravitational stratification, the value of
the Alfv\'en wave speed $C_A\equiv B_h/(4\pi\rho)^{1/2}$ may not
be very large (say, $C_A\lsim$ a few km s$^{-1}$ in the overshoot
layer suspected to be strongly magnetized).

In terms of large-scale magnetohydrodynamic (MHD) wave
propagations in a thin layer with a relatively small $C_A$ and a
relatively large azimuthal perturbation scale, the role of
subsurface magnetic field $B_h$ may be crudely assessed (Lou
1987). The presence of $B_h$ introduces a large-scale horizontal
magnetic pressure force in addition to the quasi-hydrostatic
pressure force $\rho g\eta$ caused by the surface elevation $\eta$
and thus effectively increases the surface wave speed in the form
of $c_L\sim (gD+C_A^2)^{1/2}$ where $D\lsim 500$km is an estimated
photospheric layer thickness and $g=2.7\times 10^4\hbox{ cm
s}^{-2}$ is the solar surface gravity. Replacing the surface wave
speed $c\equiv (gD)^{1/2}$ by $c_L$ here, expressions (7) and (11)
of Lou (2000b) then give increased frequencies of equatorially
trapped Kelvin- and Poincar\'e-waves (thus shorter periods than a
couple of days), whereas expressions (13) and (15) of Lou (2000b)
(or equations (3) and (4) here) show that the frequencies of
equatorially trapped Rossby-type waves remain more or less
unchanged because they are primarily determined by the solar
rotation rate $\Omega_{\odot}$ and the minor correction term
involves the speed ratio
$\Omega_{\odot}R_{\odot}/(gD+C_A^2)^{1/2}$ which is reduced
further by the presence of $B_h$. For Rossby-type waves, this
allows {\it more} nearby frequencies to be packed into the
frequency groups labeled by $m$ with different values of $n$.


\section{Other Circumstantial Evidence for Rossby-Type Waves}

While solar Rossby-type waves or $r$-modes have been proposed
along several lines of research (Papaloizou \& Pringle 1978; Wolff
1998; Lou 1987, 2000a, b), there is no direct solid evidence so
far for their detection at the solar photosphere except for a few
tantalizing cases discussed below. We intend to stimulate further
investigations of this problem both observationally and
theoretically .

The velocity correlation analysis on the data from the {\it Solar
Oscillation Investigation/Michelson Doppler Imager (SOI/MDI)}
revealed stationary ``long-lived velocity cells" (Beck et al.
1998) located along the solar equatorial zone. In terms of
horizontal surface velocities (a few meters per second) and
spatial scales ($\sim 50^{\circ}$ in longitude), it is inevitable
on theoretical ground that such large-scale ``velocity cells"
should travel in the form of Rossby-type waves relative to the Sun
at speeds slower than the solar rotation (Lou 1987, 2000a, b). An
azimuthal scale of $\sim 50^{\circ}$ may correspond to an
azimuthal wave integer $m=6$ or 7 which would give an equatorially
trapped Rossby wave period $\gsim 77$ but $\lsim 102$ days (Lou
2000b). The diagnostic approach of Beck et al. (1998) is
promising, although more sophisticated velocity correlation
analyses of the {\it SOI/MDI} data are required in order to
extract signals of large-scale, slowly drifting Rossby-types wave
patterns over the solar photosphere.

Using the solar limb data from the {\it SOI/MDI}, Kuhn et al.
(2000) reported periodic signals of 100 m high ``hills" that are
separated by an azimuthal scale of $(8.7\pm 0.6)\times 10^4$km.
Note that this surface elevation of $\sim 100$ m is some $700$
times smaller than the upper limit of 0.1'' estimated by Lou
(2000b). Uncertainties in such an estimate involves the effective
thickness $D$ of the solar photosphere and the extent of partial
cancellations in the horizontal velocity divergence
$\nabla_{\perp}\cdot\vec v_{\perp}$. The more precise result of
Kuhn et al. (2000) casts serious doubts on earlier attempted
measurements of variations in the solar diameter (e.g., Delache et
al. 1985; Ribes et al. 1987; Yoshizawa 1999). If such solar limb
elevations are indeed induced by Rossby waves with an azimuthal
wave integer $m=50$, then the corresponding equatorially trapped
Rossby wave drifts at an extremely slow speed relative to the Sun
with a period of $\sim 625$ days. It is somewhat surprising that
limb elevations caused by Rossby-type waves with smaller values of
$m$ were not detected by the same technique of Kuhn et al (2000).

Combining the data from the {\it SOI/MDI} and the {\it Global
Oscillation Network Group (GONG)}, Ulrich (2001) obtained
persistent patterns of torsional oscillations with azimuthal
structures characterized by $1<m\lsim 8$ yet without detecting
signals of quasi-periodicities. For this range of $m$ values,
periods of equatorially trapped Rossby waves should lie in the
range of $\sim 27-102$ days (Lou 2000b). It is a challenge through
a helioseismological data analysis to identify unambiguously
signatures of solar Rossby waves that involve large-scale and
long-lived structures of flow velocity and surface elevation over
the slowly rotating solar photosphere.

Ballester et al. (2002) recently examined several historical
records of solar photospheric magnetic fluxes. These records
include (1) the Mount Wilson total magnetic flux (MWTF) from 1966
to 2000 for total daily magnetic flux measured in units of
$10^{22}$Mx, (2) the Kitt Peak magnetic flux (KPMF) from 1975 to
2000 in units of $10^{22}$Mx in regions with magnetic field
strength greater than 25 G within the latitude band
S$70^{\circ}$-N$70^{\circ}$, and (3) the daily magnetic plage
strength index (MPSI) and the Mount Wilson sunspot index (MWSI)
from 1970 to 2000. They derived both periodogram and wavelet
contours. The results of their data analysis show a near 160-day
periodicity in the photospheric magnetic flux during solar cycle
21 (but not in solar cycle 22) and suggest a probable causal
relation to the well-known Rieger-type quasi-periodicities. Given
our data analyses on mid-term quasi-periodicities in CMEs and
solar X-ray flares, it is very likely that during the recent solar
maximum (cycle 23), the daily total solar magnetic flux of the
MWTF, for example, should reveal Rieger-type periodicities. The
results of Ballester et al. (2002) are consistent with the notion
that the emergence of subphotospheric magnetic flux may be
triggered or modulated quasi-periodically by large-scale Rossby
waves trapped in the solar equatorial zone (Lou 2000b).

\section{Discussion and Summary}

In terms of the interrelation between mid-term quasi-periodicities
of equatorially trapped Rossby waves and Rieger-type periodicities
in solar flare related activities, we would offer several physical
ideas here to address the interesting questions raised by
Ballester et al. (2002) regarding the mechanism proposed by Lou
(2000b). Due to the solar differential rotation, subsurface mean
magnetic fields are wrapped around the solar equator with growing
intensities. By turbulent convections and MHD instabilities,
subsurface magnetic fluxes will emerge in general and sunspots or
sunspot groups will form in particular. Magnetic activites above
the photosphere are capable of disturbing the photosphere and
sub-photospheric layers through MHD processes.

We thus have in mind a ``dynamic feedback scenario" advanced
earlier. Specifically, powerful solar flares frequently occurring
in magnetic active regions along the two usual belts across the
solar equator keep stirring the photosphere and exciting Rossby
waves that in turn either trigger or modulate the emergence of
subphotospheric magnetic flux. To sustain such a ``dynamic
feedback cycle", initiations and positive feedbacks above a
certain energetic ``threshold" are necessary. This may explain why
Rieger-type periodicities in the emergence of magnetic flux,
sunspot areas, and high-energy flares etc. are detected within a
few years around the solar maxima when solar activity levels are
high. This also implies possible changes in the most dominant
periodicities with different physical conditions for different
solar maxima.

Practically, however, it is not easy to estimate this
``threshold", because several unknown parameters and processes are
involved. Regarding possible values of $m$ and $n$ in the solar
Rossby wave theory of Lou (2000b), we already noted that for
$\sim$5 or 6 active regions in the solar disk facing us, $m$ would
be $\sim$10 or 12 around the circumference. Similarly,
equatorially trapped Rossby-type waves with $n=2$ may be more
favorably excited given the presence of two belts of sunspot
groups north and south of the equator. In the sense of triggering
magnetic avalanches, a sustained passage of Rossby-type waves can
increase {\it statistically} the chance of flare occurrence in
preexisting vulnerable magnetic complexes and thus impose grossly
quasi-periodicities to various diagnostics associated with solar
flares.

In summary, by Fourier spectrum and Morlet wavelet analyses, we
identified quasi-periods in data series of CMEs, X-ray solar
flares (class$>$M5.0), and daily averages of Ap index during the
time span from February 5, 1999 to February 10, 2003 for the solar
maximum of cycle 23. CME periods at $\sim 358\pm 38$, $\sim 272\pm
26$, and $\sim 196\pm 13$ days appear to correspond well to Ap
index periods at $\sim 364\pm 39$, $\sim 273\pm 26$, and $\sim
187\pm 12$ days, although the Ap index period of $\sim 364\pm 39$
days is somewhat less significant; we interpret these
quasi-periodic correlations in terms of CME interactions with the
Earth's magnetosphere. In addition to a significant period at
$\sim 259\pm 24$ days, X-ray flare data contains the familiar
Rieger-type periodicities shorter than $\sim 157\pm 11$ days and
so forth. Within error ranges, this $\sim 259\pm 24$-day period in
X-ray flare data may relate to CME $\sim 272\pm 26$-day and Ap
$\sim 273\pm 26$-day periods. We provide estimates for periodic
timescales of equatorially trapped Rossby-type waves based on the
work of Lou (2000a, b) and discuss a few qualitative aspects of a
``dynamic feedback cycle" to sustain such waves. A coherent
picture for solar Rossby waves is yet to come given several
preliminary results of their circumstantial detections in recent
years (Beck et al. 1998; Kuhn et al. 2000; Ulrich 2002). More
observations and theoretical investigations are needed to further
understand these mid-term quasi-periodicities in solar as well as
solar-terrestrial activities.

Finally, we emphasize the strong obvious periodicity of $187\pm
12$-days in daily averaged Ap index that can be visually picked
out in Fig. 3a. We tentatively identified a CME counterpart of
period $196\pm 13$ days as the cause. A further understanding of
this phenomenon is of great interest especially in contexts of
solar terrestrial interaction.

\section*{Acknowledgments}

This research (Y.Q.L.) was supported in part by grants from US NSF
(AST-9731623) to the University of Chicago, by the ASCI Center for
Astrophysical Thermonuclear Flashes at the University of Chicago
under Department of Energy contract B341495, by the Special Funds
for Major State Basic Science Research Projects of China, by the
Collaborative Research Fund from the NSF of China for Outstanding
Young Overseas Chinese Scholars (NSFC 10028306) at the National
Astronomical Observatories, Chinese Academy of Sciences, and by
the Yangtze Endowment from the Ministry of Education through the
Tsinghua University. Affliated institutions of Y.Q.L share this
contribution. Z.H.F. was supported in part by the NSFC grant
10243006 and the Ministry of Science and Technology of China under
grant TG1999075401.


\clearpage







\clearpage




\label{lastpage}

\end{document}